\documentclass[12pt]{article}
\font\mybb=msbm10 at 12pt
\def\bb#1{\hbox{\mybb#1}}
\def\ZZ {\bb{Z}}
\def\RR {\bb{R}}

\def\etal {{\it et al.}}

\usepackage{fullpage,epsfig,fancyheadings}

\usepackage[psamsfonts]{euscript}

\usepackage{epic}

\usepackage{amstex}

\newcommand{\beq}{\begin{equation}}
\newcommand{\eeq}{\end{equation}}

\begin{document}
\vspace*{-.6in}
\thispagestyle{empty}
\begin{flushright}
CALT-68-2019\\
hep-th/9509148\\
(revised)
\end{flushright}
\baselineskip = 20pt

\vspace{.5in}
{\Large
\begin{center}
{\bf Superstring Dualities}\footnote{Lecture
presented at the 29th International Ahrenshoop Symposium
in Buckow, Germany.}
\end{center}}
\vspace{.4in}

\begin{center}
John H. Schwarz\footnote{Work supported in part
by the U.S. Dept. of
Energy under Grant No. DE-FG03-92-ER40701.}\\
\emph{California Institute of Technology, Pasadena, CA  91125 USA}
\end{center}
\vspace{1in}

\begin{center}
\textbf{Abstract}
\end{center}
\begin{quotation}
\noindent  This talk is divided into two parts.  The first part reviews some of
the duality relationships between superstring theories.  These relationships
are interpreted as providing evidence for the existence of a unique underlying
fundamental theory.  The second part describes my recent work on the
$SL(2,\ZZ)$ duality group of the type IIB superstring theory
in ten dimensions and its
interpretation in terms of a possible theory of supermembranes in eleven
dimensions.
\end{quotation}
\vfil

\newpage

\pagenumbering{arabic} 

\section{Relationships among superstring theories}

As of ten years ago we knew five different supersymmetric string ``theories''
in ten dimensions (nine space and one time):  Types I, IIA, IIB, and two
heterotic theories.  The word ``theories'' is placed in quotation marks,
because we only had rules for characterizing a classical vacuum and for
computing quantum corrections in perturbation theory. Still, this was a major
advance over anything else on the market. Non-perturbative physics was
completely out of reach.  At the time, my colloquium lectures~\cite{schwarz}
stressed that we really didn't want five consistent quantum theories containing
gravity.  It would be much nicer if there was just one, assuming of course that
it successfully describes nature.  I suggested that some of them might turn out
to be inconsistent or some might be equivalent or
a combination of the two.  Over the past year a
great deal has been learned about non-perturbative properties of string theory,
which makes it now possible to reassess the situation.

The first two ``unifications'' to emerge were between the
two type II theories~\cite{dine}
and between the two heterotic theories~\cite{narain}.  While the IIA theory and
IIB theory in ten dimensions
are certainly different -- indeed the former is non-chiral
and the latter is chiral -- they become equivalent after compactification of
one spatial dimension on a circle.  However, this happens in a surprising way:
they are T dual.  This means that the compactification radius of the IIA
theory (call it $R_A$) is inversely proportional to the compactification radius
of the IIB theory (call it $R_B$).  The radius corresponds to the classical
value of a scalar field in nine dimensions, and the two 10D theories
correspond to two different limits of this value.  The scalar field in question
has a flat potential, so the IIA and IIB theories are smoothly connected as
boundary points of the classical moduli space of the common 9D theory.
This relationship, like most T dualities, is valid order-by-order in string
perturbation theory (though it is non-perturbative on the string world-sheet),
and is, therefore, understood very well.

The two 10D heterotic theories have gauge groups $E_8 \times E_8$ and
$O(32)$, the only possibilities allowed by anomaly cancellation.  They are
related in much the same way as the two type II theories.  Compactification to
nine dimensions gives a theory with a
17-dimensional moduli space.  One of these moduli
describes the compactification radius in the $E_8 \times E_8$ picture and
another one gives the radius
in the $O(32)$ picture.  Thus, the two 10D theories again
correspond to two distinct boundary points of the classical moduli
space for nine dimensions.

We have now reduced the number of ``theories'' to three -- type I, type II, and
heterotic.  To do better than this requires going beyond string perturbation
theory, and is necessarily more speculative.
After reviewing the evidence, I'll present my interpretation
of its significance.

The type II theory is related to the heterotic theory by ``string-string
duality''~\cite{hulla,duffsix,witten,sensix,harvey}.
It has various manifestations, but the basic one is
described in six dimensions.  The conjecture asserts that the type IIA theory
compactified on the K3 manifold is non-perturbatively equivalent to the
heterotic theory compactified on a four-torus.  Both theories have the same
massless sectors and the same $O(4,20;\ZZ)$ discrete group of symmetries.
The situation is much the same as before. Namely, the 10D theories
correspond to different boundary points in the 80-dimensional moduli
space of the common 6D theory.
Moreover, the common low-energy effective supergravity theory gives rise to
both kinds of strings as soliton-like solutions.  These arguments, as well as a
variety of others, make a compelling case.  This identification is remarkable,
because weak coupling is mapped to strong coupling, and therefore
perturbative results in one description correspond to non-perturbative
results in the other.

All that now remains is to relate the type I superstring theory
to the others.  This appears very challenging, because type I superstrings
are unoriented open and
closed strings, whereas type II and heterotic strings are
oriented closed strings.
However, these strings are used to define perturbation
expansions, and the identifications are supposed to be
non-perturbative, so this may not be as big an obstacle
as we used to believe. One
could imagine that heterotic or type II strings would arise as
solitons of the type I theory and vice-versa. Indeed,
the low-energy supergravity theories obtained from the 10D
$O(32)$ type I and heterotic strings are the same, but the correspondence
implies that the dilaton field obtained from the one
string corresponds to the negative of the dilaton obtained from the other,
which means that the
coupling constants are inversely proportional~\cite{witten}.
This motivated the conjecture that the string theories
are really the same, and
that the perturbative expansion of one corresponds to the
strong coupling expansion of the other
(just as in the case of the 6D string-string duality described above).
While it is understood how to obtain the
heterotic string as a soliton solution of the low-energy supergravity
theory~\cite{dabholkar10,hullb},
a type I string solution would involve additional subtleties and has not
yet been constructed.  Curiously, there is no candidate for an analogous dual
string description of the $E_8 \times E_8$ heterotic string in ten dimensions.

We have now connected all five of the 10D superstring theories,
but there is one more ``theory'' that should be mentioned, namely 11D
supergravity.  Unlike the string theories,
it is not perturbatively renormalizable, but it might exist as a quantum
theory nonperturbatively.
Townsend \cite{townsend11} and Witten \cite{witten} have
pointed out that the 10D type IIA theory is
actually eleven dimensional!  The IIA theory has a hidden circular
eleventh dimension, whose radius scales as the two-thirds power of the string
coupling constant.  Thus, in perturbation theory the IIA theory looks ten
dimensional, but in the strong coupling limit one obtains full
11D Poincar\'e symmetry described at low energy
by an effective 11D supergravity theory.

Having reviewed the relevant facts, let us now contemplate their implications.
It seems to me that these duality relations imply that the
various different string ``theories'' should be viewed as
recipes for finding classical solutions (and their perturbative quantum
improvements) of a single underlying theory.  What this underlying
theory should be is quite uncertain at this time, but it may be unlike anything
we've seen before.  It needs to be able to give rise to all the solutions
obtainable by any of the known recipes and account for all of their duality
symmetries and relationships.  It might also have
additional solutions that are not obtainable by any of the currently known
recipes. The remarkable role of
duality symmetries and their geometrically non-intuitive
implications suggest to me that the theory might look
very algebraic in structure without evident
geometric properties, so that no space-time manifold is evident in its
formulation.  In this case, the existence of space-time
would have to emerge as a property of a class of
solutions. Other solutions might not have any such interpretation.
Even stringy one-dimensional structures might not play a more
prominent role than other $p$-branes in the underlying
theory~\cite{townsendb,becker},
in which case the subject will require a new name. One line of
inquiry that may be helpful for inventing the theory
is to focus on determining the complete group of
duality symmetries.  Assuming this is a well-defined notion
(so that the group that appears is
not just an artifact of formalism), the duality group could turn
out to be some very large discrete subgroup of a hyperbolic Lie algebra such as
$E_{10}$ or the monster Lie algebra.
In the past, symmetries have been a useful
guide to dynamics.  Maybe that will turn out to be the case once again.

On the other hand, a radically different type of description,
which is much more geometrical and less algebraic,
could turn out to be correct. One such possibility
is a theory based on fundamental supermembranes in eleven dimensions
\cite{bergshoeff,duffstelle}.
This proposal used to look very unattractive to me, because of
its bad perturbative quantum behavior and its lack of chirality.
Now, I am beginning to take it much more seriously, because there are
remarkable heuristic arguments for how various string theories could
be deduced from such a starting point. The next section describes
how it can lead to the {\it chiral} IIB theory in ten dimensions.
Most of the results that follow
have been presented previously in ref. \cite{twob}, but one new result --
an interpretation of type IIB superstrings as wrapped supermembranes --
is given at the end of the section.

\section{Type IIB Superstrings}

Among the various conjectured duality symmetries of superstring theories, the
proposed $SL(2,\ZZ)$ symmetry of the type IIB superstring theory in ten
dimensions is especially interesting~\cite{hulla,greena}. Like the $SL(2,\ZZ)$
S duality of the $N=4$
4D heterotic string~\cite{font,sen}, it relates weak and
strong coupling.  However, unlike the heterotic example in which the symmetry
relates particles carrying electric and magnetic charges of the same gauge
field,
the IIB duality relates strings carrying electric charges of two different
gauge
fields.  In this respect it is more like a T duality \cite{giveon}.  Combined
with ordinary T dualities, the IIB $SL(2,\ZZ)$ duality implies the complete U
duality symmetry of toroidally compactified type II strings in dimensions
$D < 10$ \cite{hulla,witten}.

The $SL(2,\ZZ)$ duality of the IIB theory will be explored here by considering
string-like (or `one-brane')
solutions of the 10D IIB supergravity theory.  It will be
argued that there is an infinite family of such solutions
forming an $SL(2,\ZZ)$
multiplet. (This possibility was hinted at in section 5 of Ref. \cite{harvey}.)
Once these string solutions have been constructed, we will consider
compactification on a circle and compare the resulting
9D spectrum with that of
11D supergravity compactified on a two-torus.
The conclusion will be that the $SL(2,\ZZ)$ duality
group of the IIB theory
{\it in ten dimensions} corresponds precisely to the modular
group of the torus,\footnote{This conclusion has been reached independently
by Aspinwall \cite{aspinwall}. The 11D origin of the
$SL(11-d,Z)$ subgroup of the U duality group
of type II string theory in $d$ dimensions
is explained in Witten's paper \cite{witten}.
See also Appendix B of Ref. \cite{bergshoeffc}.}
and that all type II superstrings can be interpreted
as wrapped supermembranes of 11D supergravity.

All 10D supergravity theories contain the following terms in common
\begin{equation}
{S}_0 = {1\over 2\kappa^2}
\int d^{10} x \sqrt{-g} \left(R - {1\over 2}
(\partial \phi)^2 - {1\over 12} e^{-\phi} H^2\right),
\end{equation}
where $H$ is a three-form field strength $(H = dB)$, and $\phi$ is the dilaton.
Moreover, in each case, a solution to
the classical equations of motion derived
from $S_0$ can be regarded as a solution
of the complete supergravity theory with
all other fields set equal to zero.  A macroscopic string-like solution,
which was identified with the heterotic string,
was constructed by Dabholkar \etal \, \cite{dabholkara}  (This was
generalized to $p$-branes in Ref.~\cite{horowitz}.)
Restricted to ten dimensions, it is given by
\begin{equation}
ds^2 = A^{-3/4} [-dt^2 + (d {x}^1)^2] + A^{1/4} d {{\bf x}} \cdot
d{{\bf x}}, \label{metric}
\end{equation}
\begin{equation}
B_{01} = e^{2\phi} = A^{-1}, \label{bvalue}
\end{equation}
where
\begin{equation}
A = 1 + {Q\over 3r^6} , \label{Aeqn}
\end{equation}
${\bf x} = ({ x}^2, {x}^3, ... , {x}^9)$, ${{\bf x}} \cdot
{{\bf x}} = r^2 = \delta_{ij} {x}^i {x}^j$, and $Q$ is the
$H$ electric charge carried by the string.  (Recall that
the electric charge of a $(p +
2)$-form field strength is carried by a $p$-brane.)  Strictly
speaking, the $S_0$ equations are not satisfied at $r = 0$, the string
location, because $\nabla^2 A$ has a delta-function singularity there.  In Ref.
{}~\cite{dabholkara} it was proposed that this could be fixed by
coupling to a string source, which means considering $S
= S_0 + S_\sigma$ instead, where
\begin{equation}
S_\sigma = - {T\over 2} \int d^2 \sigma (\partial^\alpha X^\mu
\partial_\alpha X^\nu G_{\mu\nu} + \ldots),
\end{equation}
$T$ is the string tension, and
$G_{\mu\nu} = e^{\phi/2} g_{\mu\nu}$
is the string metric.  From the coefficient of the delta function one can
deduce the relation $Q = {{\kappa}^2 T/ \omega_7}$,
where $\omega_7 = {1\over 3}
\pi^4$ is the volume of $S^7$.

The type IIB theory has two three-form field strengths $H^{(i)} = dB^{(i)}$,
$i = 1,2$~\cite{greenb}.  $H^{(1)}$ belongs to the NS--NS sector and can be
identified with $H$ in the preceding discussion.  $H^{(2)}$ belongs to the
R--R sector and does not couple to the (usual) string world sheet.  In
addition, the type IIB theory has two scalar fields, which can be combined into
a complex field $\lambda = \chi + ie^{-\phi}$.  The dilaton $\phi$ is in the
NS--NS sector and can be identified with $\phi$ in the preceding discussion,
while $\chi$ belongs to the R--R sector.  The other bose fields are the metric
$g_{\mu\nu}$ and a self-dual five-form field strength $F_5$.  The five-form
field strength will be set to zero, since
the corresponding charges are carried by a self-dual
three-brane, whereas the focus here is on charges carried by strings.  Once we
set $F_5 = 0$, it is possible to write down a covariant action that gives the
desired equations of motion~\cite{hullb}:
\begin{equation}
S = {1\over 2{\kappa}^2}
\int d^{10} { x} \sqrt{-g} (R + {1\over 4} tr (\partial
\mathcal{M} \partial \mathcal{M}^{-1}) - {1\over 12} H^T \mathcal{M} H).
\end{equation}
Here we have combined $H^{(1)}$ and $H^{(2)}$ into a two-component vector $H =
dB$, and introduced the symmetric $SL(2,\RR)$ matrix
\begin{equation}
\mathcal{M} = e^\phi
\left( \begin{array}{cc} | \lambda |^2 & \chi\\
\chi & 1 \end{array} \right).
\end{equation}
This action has manifest invariance under the global $SL(2,\RR)$ transformation
\begin{equation}
\mathcal{M} \rightarrow \Lambda \mathcal{M} \Lambda^T, \quad B \rightarrow
(\Lambda^T)^{-1} B.
\end{equation}
The choice $\Lambda = \left(\begin{array}{cc} a & b\\ c & d \end{array}
\right)$ corresponds to
\begin{equation}
\lambda \rightarrow {a \lambda + b\over c\lambda + d}, \qquad
B^{(1)} \rightarrow d B^{(1)} - c B^{(2)}, \qquad
B^{(2)} \rightarrow a B^{(2)} - b B^{(1)}.
\end{equation}

Given the symmetry of this system, it is clearly artificial to only consider
solutions carrying $H^{(1)}$ electric charge and not $H^{(2)}$ electric charge.
Measured in units of $Q$, we will consider solutions carrying charges $(q_1,
q_2)$. Since there exist five-brane solutions carrying magnetic
$H$ charge~\cite{horowitz},
the generalized Dirac quantization condition~ \cite{nepomechie}
implies that $q_1$ and $q_2$ must be integers.  Moreover, $q_1$ and $q_2$
should be relatively prime,  since otherwise the solution is neutrally stable
against decomposing into a multiple string solution --- the number of strings
being given by the common divisor.\footnote{This is the same counting rule
that was required in a different context in Ref. \cite{strominger}.
We will show that it leads to sensible degeneracies after compactification
on a circle. As was pointed out in Ref. \cite{strominger},
a different rule is sometimes appropriate in other situations.}
Also, the $(q_1, q_2)$ string and the
$(-q_1, -q_2)$ string are related by orientation reversal $({x}^1
\rightarrow -{x}^1)$.
A complete description of string solutions requires specifying
the vacuum in which they
reside.  In the IIB theory this means choosing
the asymptotic value of $\lambda$ as
$r \rightarrow \infty$, denoted by $\lambda_0$.  The simplest choice
is $\lambda_0 = i$, which corresponds
to $\chi_0 = \phi_0 = 0$. The $(1,0)$ string in this
background is given by the solution in eqs. (\ref{metric}) and (\ref{bvalue}).
By applying an appropriate $SL(2,\RR)$
transformation to that solution, we can obtain
the solution describing the $(q_1, q_2)$ string for arbitrary $\lambda_0$.

The general solution describing a $(q_1, q_2)$ string in the
$\lambda_0$ vacuum found in ref. \cite{twob} is
\begin{equation}
\quad ds^2 = A_q^{-3/4} [-dt^2 + (d {x}^1)^2] + A_q^{1/4} d {{\bf x}} \cdot
d{{\bf x}},
\end{equation}
\begin{equation}
B^{(i)}_{01} = q_i \Delta_q^{-1/2} A_q^{-1},\quad \qquad \qquad \qquad \qquad
\end{equation}
\begin{equation}
\lambda = {i(q_2 \chi_0 + q_1 |\lambda_0|^2) A_q^{1/2} - q_2 e^{-\phi_{0}}\over
i(q_1 \chi_0 + q_2) A_q^{1/2} + q_1 e^{-\phi_{0}}},
\end{equation}
where\footnote{The original preprint version of this paper, and
also of Ref. \cite{twob}, had an error in eq. (\ref{Delta}).}
\begin{equation}
\Delta_q = (q_1\ q_2) \mathcal{M}_0^{-1} \left(\begin{array}{c}
q_1\\  q_2 \end{array} \right) = e^{\phi_{0}} (q_2 \chi_0 - q_1)^2 +
e^{-\phi_{0}} q_2^2, \label{Delta}
\end{equation}
\begin{equation}
A_q(r) = 1 + \Delta_q^{1/2} {Q \over 3r^6}.
\end{equation}
These equations describe an $SL(2,\ZZ)$ family of
type IIB macroscopic strings
carrying $H$ charges $(q_1, q_2)$ and
an $SL(2,\ZZ)$ covariant spectrum of tensions given by
\begin{equation}
T_q = \Delta_q^{1/2} T. \label{tension}
\end{equation}
For generic values of $\lambda_0$ one of these tensions is smallest.  However,
for special values of $\lambda_0$ there are degeneracies.  For example,
$T_{1,0}
= T_{0,1}$ whenever $|\lambda_0| = 1$.  (More generally, $T_{q_1,q_2}
=T_{q_2,q_1}$ in this case.)
Also, $T_{0,1} = T_{1,1}$ whenever $\chi_0
= {1\over 2}$.   (More generally, $T_{q_1,q_2}
=T_{q_2 -q_1, q_2}$ in this case.)
Combining these, we find a three-fold degeneracy $T_{1,0} =
T_{0,1} = T_{1,1}$ for the special choice $\lambda_0 = e^{\pi i/3}$.

Although we have only constructed infinite straight macroscopic strings, there
must be an infinite family of little loopy strings whose spectrum of
excitations can be analyzed in the usual way.  Thus, in ten dimensions
each of the
$(q_1, q_2)$ strings has a perturbative spectrum given by $M^2 = 4\pi
T_q (N_L + N_R)$, where $N_L$ and $N_R$ are made from oscillators in the usual
way.  Each of the strings has the same massless sector --- the IIB
supergravity multiplet --- in common.  The excited states are presumably
distinct, with the excited levels of one string representing states that are
non-perturbative from the viewpoint of any of the other strings.  Of course,
the formula for $M^2$ gives the free-particle spectrum only, which
is not meaningful for
two different strings at the same time, so comparisons
of massive levels are only
qualitative.  In ten dimensions, the only states in short supersymmetry
multiplets, for which we have good control of the corrections, are
those
of the supergravity multiplet itself.  We now turn to the theory compactified
to nine dimensions, because much more of the spectrum is under precise control
in that case.

Consider the $(q_1, q_2)$ IIB string compactified on a circle of radius $R_B$.
Then the resulting perturbative spectrum of this string  has 9D
masses given by
\begin{equation}
M_B^2 = \left({m\over R_B}\right)^2 + (2\pi R_B n T_q)^2 + 4\pi T_q (N_L +
N_R), \label{bmass}
\end{equation}
where $m$ is the Kaluza--Klein excitation number (discrete momentum) and $n$ is
the winding number, as usual.  Level-matching gives the condition
$N_R - N_L = mn$.
Short multiplets (which saturate a BPS bound) have $N_R = 0$ or $N_L =
0$. (Ones with $N_L=N_R=0$ are `ultrashort'.)
Taking $N_L =0$ gives $M_B^2 = ( 2\pi R_B n T_q +{m/ R_B})^2$
and a rich spectrum controlled by $N_R = mn$.
The masses of these states should be exact, and they should be stable in
the exact theory.  Note that
\begin{equation}
n^2 T_q^2 = [\ell_2^2 + e^{2\phi_{0}} (\ell_2 \chi_0 - \ell_1)^2]e^{-\phi_{0}}
T^2,
\end{equation}
where $\ell_1 = nq_1$ and $\ell_2 = nq_2$.  Any pair of integers $(\ell_1,
\ell_2)$ uniquely determines $n$ and $(q_1, q_2)$ up to an irrelevant sign
ambiguity.  Winding a $(-q_1, -q_2)$ string $-n$ times is the same thing as
winding a $(q_1, q_2)$ string $n$ times.  Thus the pair of integers $(\ell_1,
\ell_2)$ occurs exactly once, with the tension of the string determined by the
corresponding pair $(q_1, q_2)$.

Since the IIB theory compactified on a circle is equivalent to the
IIA theory compactified on a circle of reciprocal radius, and
the IIA theory corresponds to 11D supergravity
compactified on a circle, there should be a correspondence between
the IIB theory compactified on a circle and 11D supergravity
compactified on a torus.\footnote{A detailed comparison of the 9D fields
and dualities obtained by compactifying the 11D, IIA, and IIB
supergravity theories is
given in Ref. ~\cite{bergshoeffa}.}
Therefore, let us consider compactification of 11D supergravity on
a torus with modular parameter $\tau = \tau_1 + i\tau_2$.  The
Kaluza--Klein modes on this torus are described by wave functions
\begin{equation}
\psi_{\ell_{1},\ell_{2}} (x,y) \sim \exp \left\{{i\over R_{11}} \left[{x}
\ell_2 - {1\over \tau_2} y (\ell_2 \tau_1 - \ell_1)\right]\right\} \quad
\ell_1,\ell_2 \in \ZZ.
\end{equation}
Letting $z =  (x + iy){/2\pi R_{11}}$,
$\psi_{\ell_{1},\ell_{2}}$ is
evidently invariant under $z \rightarrow z + 1$ and $z \rightarrow z + \tau$.
The contribution to the 9D mass-squared is given by the
eigenvalue of $p_x^2 + p_y^2 = - \partial_x^2 - \partial_y^2$.  Let us try to
take the supermembrane idea ~\cite{bergshoeff,duffstelle,townsend11}
seriously, and suppose that
it has
tension (mass/unit area) $T_{11}$.  Wrapping it so that it covers
the torus $m$ times gives a contribution to the mass-squared of $(m A_{11}
T_{11})^2$. (Different maps giving the same $m$ are identified.)
The area of the torus in the 11D metric is
$A_{11} = (2 \pi R_{11})^2 \tau_2$.  Therefore, states with wrapping number $m$
and Kaluza--Klein excitations $(\ell_1, \ell_2)$ have 9D
mass-squared (in the 11D metric)
\begin{equation}
M_{11}^2 = \Big(m (2 \pi R_{11})^2 \tau_2 T_{11}\Big)^2
+ {1\over R_{11}^2} \Big(\ell_2^2 +
{1\over \tau_2^2} (\ell_2 \tau_1 - \ell_1)^2 \Big) + \dots~~,
\label{elevenmass}
\end{equation}
where the dots represent membrane excitations, which we do not know how to
compute.  This is to be compared to eq. (\ref{bmass}) for $M_B^2$, allowing
$M_{11} = \beta M_B$, since they are measured in different metrics.  Agreement
of the formulas for the masses of BPS saturated states
is only possible if the vacuum modulus $\lambda_0$ of the IIB
theory is identified with the modular parameter $\tau$ of the torus.  Since
$SL(2, \ZZ)$ is the modular group of the torus, this provides strong evidence
that it should also be the duality group of the IIB string.  In addition, the
identification $M_{11} = \beta M_B$ gives
\begin{equation}
R_B^{-2} = T T_{11} A_{11}^{3/2}, \label{rbeqn}
\end{equation}
\begin{equation}
\beta^2 = 2\pi R_{11} e^{-\phi_0 /2}  T_{11}/T.
\end{equation}
These identifications imply predictions for the spectrum of membrane
excitations -- at least those that give short supermultiplets.

It is also interesting to explore how the type IIA string fits into the story.
One can identify $r_{11}= R_{11}\tau_2$ as the radius of the circle that
takes 11D supergravity to the 10D IIA theory, and $R_{11}$
as the radius of circle taking the 10D IIA theory to nine dimensions.
Type IIA and IIB superstrings in
nine dimensions are usually considered to be
related by an $R_B  \sim 1/R_A$ duality of two circles, as has
already been mentioned.
The identification of eqs. (\ref{bmass}) and (\ref{elevenmass})
can be interpreted to mean that wrappings of the supermembrane on the
torus in the 11D theory
correspond to Kaluza--Klein excitations of the circle of the IIB theory,
and windings of an $SL(2,\ZZ)$ family of type IIB strings on the circle
correspond to Kaluza--Klein excitations of the torus. So the duality
is really between a torus and a circle rather than between
two circles. Equation (\ref{rbeqn}) tells us that $R_B \sim A_{11}^{-3/4}$.

In Ref. ~\cite{witten}, Witten showed that
$r_{11} \sim e^{2 \phi_{A}/3}$, where
$\phi_A$ is the vev of the dilaton field in
the $D = 10$ IIA theory, by comparing the masses of
IIA states in ten dimensions
that saturate a BPS bound to those of Kaluza--Klein
excitations of 11D supergravity compactified on a circle of radius $r_{11}$.
An alternative
approach is to consider wrapping the 11D supermembrane on a circle of radius
$r_{11}$ to give a type IIA string with tension
$2\pi r_{11} T_{11}$ \cite{duffa}.
This tension is measured in
units of the 11D metric $g^{(11)}$.
It can be converted to the
IIA string metric $g_{A}^{(10)}$ using
$g^{(11)} = e^{-2\phi_{A}/3} g_{A}^{(10)}$.
Denoting the IIA string tension in the IIA string metric by
$T_{A}$, we deduce that
\begin{equation}
T_{A} = 2 \pi r_{11} T_{11} e^{-2\phi_{A}/3}.
\end{equation}
Since $T_A$ and $T_{11}$ are constants independent of $r_{11}$ and $\phi_A$,
we have confirmed that $r_{11} \sim e^{2\phi_{A}/3}$ and even
determined the constant of proportionality.

Type IIB superstrings also have a simple
supermembrane interpretation. To see this,
let us begin with the 11D theory compactified on a two-torus
and consider a
toroidal supermembrane with one of its cycles
mapped onto a $(q_1, q_2)$ cycle of the spatial torus.
The other membrane cycle describes the
resulting 9D string. The integers $q_1$ and $q_2$
are taken to be relatively prime for the same reasons as before.
Now suppose the membrane shrinks to the shortest cycle in the
$(q_1, q_2)$ homology class. Then it can be represented as
a straight line from $z=0$ to $z = q_1 \tau + q_2$,
which has length $L_q = 2 \pi R_{11} |q_1 \tau + q_2| $.
The resulting 9D string has tension (in the 11D metric)
$T_q^{(11)} = L_q T_{11}$. Converting to the IIB metric by
setting $T_q = \beta^{-2} T_q^{(11)}$
precisely reproduces the previous
result $T_q = \Delta_q^{1/2} T$. This means that
in nine dimensions the
entire $SL(2, \ZZ )$ family of type
II superstrings can be regarded as different wrappings of a
unique supermembrane! Equation (\ref{rbeqn}) implies that the
10D IIB superstring
theory can be recovered by taking the limit in which the torus
is shrunk to a point ($A_{11} \to 0$) while holding the modulus
$\tau = \lambda_0$ fixed. Remarkably, the tensions
$T_q$ are all independent of $A_{11}$, and therefore they are
finite in the limit.

One can also identify the IIA theory in ten dimensions
as a decompactification of the toroidally compactified supermembrane theory
described previously. For this purpose, we let $R_{11}\to \infty$
while holding $r_{11} = R_{11} \tau_2$ fixed.
In this limit the only
$(q_1, q_2)$ string whose tension remains finite is the $(0,1)$
string, whose tension in the 11D metric is $2\pi r_{11} T_{11}$.
(The other strings become zero-branes/black holes in the limit.)
This limit implies that
$\tau$ is outside the usual $SL(2,\ZZ)$ fundamental region,
but it can be mapped into
the usual fundamental region by the $SL(2,\ZZ)$ transformation
$\tau \to - 1/\tau$. This turns the $(0,1)$ string into the
$(1,0)$ string, and thus the $(1,0)$
string is identified as the (unique) fundamental IIA
string. This resolves the puzzle of why a
supermembrane can wrap around a two-torus any number of times, but it
can wrap around a circle only once, at least if one
accepts that $q_1$ and $q_2$ should be relatively prime.

To summarize, there is an infinite family of type IIB superstrings
in ten dimensions labeled by a
pair of relatively prime integers, which correspond to their $H$ charges.  Any
one of the strings can be regarded as fundamental with the rest describing
non-perturbative aspects of the theory. This family of strings has tensions
given by the $SL(2,\ZZ)$ covariant expression in eqs. (\ref{Delta})
and (\ref{tension}).
Compactifying on a circle to nine dimensions
and identifying with 11D supergravity
compactified on a two-torus requires
equating the modulus $\lambda_0$ of the IIB
theory and the modular parameter $\tau$ of the torus, which is strong evidence
for $SL(2,\ZZ)$ duality.  Other aspects of the spectrum are consistent with a
supermembrane interpretation.

\section{Conclusion}

We have learned the following:
\begin{enumerate}

\item  There are many conjectured non-perturbative dualities,
all of which seem to be true.

\item  Type IIB superstring theory
in ten dimensions has an infinite multiplet of strings
with an $SL(2, \ZZ)$ covariant spectrum of tensions.
At least heuristically, the strings can be described as
different wrappings of a supermembrane on a two-torus. This suggests that
supermembranes might be more than just a heuristic tool.

\item  There ought to be a completely unique underlying ``superstring theory.''
Whether it is based on something geometrical (like supermembranes)
or something completely different is still not known.  In any case,
finding it would be a landmark in human intellectual history.

\end{enumerate}

I wish to thank the conference organizers, especially Dieter L\"ust and
Gerhard Weigt, for their hospitality.

\vfill\eject

\end{document}